\documentclass[12pt]{article}
\usepackage{amssymb,epsfig}
\newcommand{\be}{\begin{eqnarray}}
\newcommand{\ee}{\end{eqnarray}}
\newcommand{\bea}{\begin{eqnarray*}}
\newcommand{\eea}{\end{eqnarray*}}
\newcommand{\pa}{\partial}
\renewcommand{\d}{{\rm d}}
\newcommand{\D}{{\rm D}}

\title{\bf Noncommutative Korteweg-de-Vries equation}

\date{  }

\author{A. Dimakis \\ Department of Mathematics, University of the Aegean \\
        GR-83200 Karlovasi, Samos, Greece \\ dimakis@aegean.gr
        \\[2ex]
        F. M\"uller-Hoissen \\ Max-Planck-Institut f\"ur Str\"omungsforschung \\
        Bunsenstrasse 10, D-37073 G\"ottingen, Germany \\
        fmuelle@gwdg.de }

\begin{document}

\renewcommand{\theequation} {\arabic{section}.\arabic{equation}}

\maketitle

\begin{abstract}
We construct a deformation quantized version (ncKdV) of the KdV equation 
which possesses an infinite set of conserved densities. Solutions of the 
ncKdV are obtained from solutions of the KdV equation via a kind of Seiberg-Witten 
map. The ncKdV is related to a modified ncKdV equation by a noncommutative Miura 
transformation.
\end{abstract}

\section{Introduction}
\setcounter{equation}{0}
 Field theories on noncommutative spaces and more specifically Moyal deformed 
space-times, gained a lot of interest recently because of the appearance of 
such theories as certain limits of string, D-brane and M theory (see 
\cite{Seib+Witt99} and the references cited there). 
In this letter we apply deformation quantization \cite{dq} to a classical 
integrable model, the KdV equation. 
The passage from commutative to noncommutative space-time is achieved by 
replacing the ordinary commutative product in the space 
of smooth functions on $\mathbb{R}^2$ with coordinates $t,x$ by the 
noncommutative associative (Moyal) $\ast$-product \cite{dq} which is 
defined by
\be
   f \ast h = {\bf m} \circ e^{\vartheta P/2} (f \otimes h)   \label{ast}
\ee
where $\vartheta$ is a real or imaginary constant and 
\be
   {\bf m} (f \otimes h) = f \, h \, , \qquad
   P = \pa_t \otimes \pa_x - \pa_x \otimes \pa_t  \; .
\ee
\vskip.1cm

An essential ingredient of our deformation of the KdV equation is the
concept of a {\em bicomplex}. This is an ${\mathbb{N}}_0$-graded linear 
space (over ${\mathbb{R}}$ or ${\mathbb{C}}$) $M = \bigoplus_{r \geq 0} M^r$ 
together with two linear maps $\d , \delta \, : \, M^r \rightarrow M^{r+1}$ 
satisfying
\be
   \d^2 = 0 \, , \qquad  \delta^2 = 0 \, , \qquad  
   \d \, \delta + \delta \, \d = 0 \; .   \label{bicomplex_cond}
\ee 
Associated with a bicomplex is the {\em linear equation} 
\be
    \delta \chi = \lambda \, \d \, \chi  
               \label{bc-linear}
\ee
where $\chi \in M^0$.\footnote{See \cite{DMH00a,DMH00b} for further details and 
several generalizations.} 
Let us assume that it admits a (non-trivial) solution as a (formal) power series 
$\chi = \sum_{m \geq 0} \lambda^m \chi^{(m)}$ in $\lambda$.
The linear equation then leads to 
\be
   \delta \chi^{(0)} = 0 \, , \quad
   \delta \chi^{(m)} = \d \chi^{(m-1)} \, , \quad m=1, \ldots, \infty \; .
\ee 
As a consequence, $J^{(m+1)} = \d \chi^{(m)}$ ($m=0, \ldots, \infty$) are 
$\delta$-exact. These elements of $M^1$ should be regarded as generalized conserved 
currents (see \cite{DMH00a,DMH00b}).
\vskip.1cm

Starting with a certain trivial bicomplex, a dressing (in the sense of 
\cite{DMH00b}) which involves the $\ast$-product results in bicomplex 
equations which are equivalent to a deformed KdV equation.\footnote{See 
also \cite{DMH99} for a bicomplex treatment of the classical KdV equation.} 
As a consequence of the underlying bicomplex structure, it shares with the 
classical equation the property of possessing an infinite set of conservation 
laws which is a characteristic property of soliton equations. The same 
procedure has been applied in \cite{DMH00c,DMH00d} to obtain ``quantized" 
versions of other integrable models.
\vskip.1cm

In section 2 we derive the ncKdV equation and demonstrate the existence of 
an infinite set of conservation laws.
Section 3 shows how solutions of the KdV equation determine solutions 
of the ncKdV equation. This is similar to the Seiberg-Witten map 
between commutative and noncommutative gauge field theories \cite{Seib+Witt99}. 
In particular, we find that the one-soliton KdV solution is also an exact 
solution of the ncKdV equation. For the KdV two-soliton solution, however, 
there are corrections involving the deformation parameter. 
Several familiar classical structures generalize to the noncommutative framework. 
This concerns in particular the relation between the KdV and the modified KdV 
equation via the Miura transformation, as shown in section 4. 
 Finally, section 5 contains some conclusions.

\section{The KdV equation in noncommutative space-time}
\setcounter{equation}{0}
We choose the bicomplex space as $M = M^0 \otimes \Lambda$ where 
$M^0 = C^\infty({\mathbb{R}^2})$ and $\Lambda = \bigoplus_{r=0}^2 \Lambda^r$ 
is the exterior algebra of a $2$-dimensional vector space with basis $\tau, \xi$ 
of $\Lambda^1$ (so that $\tau^2 = \xi^2 = \tau \, \xi + \xi \, \tau = 0$). 
It is then sufficient to define bicomplex maps $\d$ and $\delta$ on $M^0$ 
since by linearity and $\d (f \, \tau + h \, \xi) 
= (\d f) \, \tau + (\d h) \, \xi$ (and correspondingly for $\delta$) for smooth 
functions $f,h$ they extend to the whole of $M$.
\vskip.1cm

Let us start with the trivial bicomplex which is given by
\be
     \d f &=& - f_{xx} \, \xi + ( f_t + 4 \, f_{xxx} ) \, \tau    \\
 \delta f &=& f_x \, \xi - 3 \, f_{xx} \, \tau 
\ee
where a subscript denotes partial differentiation, e.g., $f_{xx} = \pa_x^2 f$ 
with $\pa_x = \pa/\pa x$. 
Now we apply a dressing \cite{DMH00b} to the bicomplex map $\d$,
\be
  \D f &=& \d f + \delta (\phi \ast f) - \phi \ast \delta f  \nonumber \\
       &=& (-f_{xx} + u \ast f) \, \xi + (f_t + 4 \, f_{xxx} - 6 \, u \ast f_x 
           - 3 \, u_x \ast f) \, \tau
\ee 
where $\phi$ is a function and $u = \phi_x$. Here we used that the partial 
derivatives $\pa_x$ and $\pa_t$ are derivations with respect to the $\ast$-product. 
The only nontrivial bicomplex equation is now $\D^2 = 0$ which is equivalent to the 
{\em noncommutative KdV} (ncKdV) equation 
\be
    u_t + u_{xxx} - 3 \, (u \ast u_x + u_x \ast u) = 0  \; . \label{ncKdV}
\ee
\vskip.1cm

The linear system $\delta \chi = \lambda \, \D \chi$ associated with the 
ncKdV equation reads
\be
   \chi_x = \lambda \, (u \ast \chi - \chi_{xx}) \, ,  \qquad                
   \chi_{xx} = - {1 \over 3} \, \lambda \, (\chi_t + 4 \, \chi_{xxx} 
               - 6 \, u \ast \chi_x - 3 \, u_x \ast \chi) \; .
              \label{chi-eq}
\ee
Now we introduce functions $p$ and $q$ such that
\be
   \chi_t = - \lambda \, p \ast \chi \, , \qquad
   \chi_x = \lambda \, q \ast \chi     \label{def_pq}
\ee
assuming that $\chi$ is $\ast$-invertible. From $\chi_{tx} = \chi_{xt}$ 
and (\ref{def_pq}) we find
\be
   q_t + p_x - \lambda \, (q \ast p - p \ast q) = 0 \; .  
                                    \label{cl}
\ee
Using the product \cite{Stra97,DMH00d} defined by 
\be
   f \diamond h 
 = {\bf m} \circ \frac{\sinh(\vartheta P/2)}{\vartheta P/2}(f \otimes h)
             \label{diamond}
\ee
this can be written in the form of a conservation law as follows,
\be
    w_t + (p + \lambda \, \vartheta \, q \diamond p_t)_x = 0
\ee
where
\be
    w = q - \lambda \, \vartheta \, q \diamond p_x   \; .
           \label{w}
\ee
In terms of $p$ and $q$, the equations (\ref{chi-eq}) take the form
\be
   q &=& u - \lambda \, q_x - \lambda^2 q \ast q 
         \label{q-eq}  \\
   p &=& q_{xx} + 3 \, q \ast q - 6 \, u \ast q 
         + \lambda \, (5 \, q_x \ast q + q \ast q_x)
         + 4 \, \lambda^2 \, q \ast q \ast q \nonumber \\
     &=& q_{xx} - q \ast q - 2 \, u \ast q + \lambda \, (q \ast q)_x
          \label{p-eq}
\ee
where (\ref{q-eq}) has been used twice to simplify the expression for $p$. 
Let us expand $p$ and $q$ into power series in $\lambda$,
\be
    p = \sum_{m=0}^\infty \lambda^m \, p^{(m)} \, , \qquad
    q = \sum_{m=0}^\infty \lambda^m \, q^{(m)} \; .
\ee
Then (\ref{q-eq}) leads to
\be
    q^{(0)} = u     \, , \qquad
    q^{(1)} = - u_x
\ee
and 
\be
   q^{(m)} = - q^{(m-1)}_x - \sum_{k=0}^{m-2} q^{(k)} \ast q^{(m-2-k)} 
\ee
for $m>1$. From (\ref{p-eq}) we get
\be
    p^{(0)} = u_{xx} - 3 \, u \ast u  \, , \qquad
    p^{(1)} = - u_{xxx} + 2 \, u_x \ast u + 4 \, u \ast u_x 
\ee
and 
\be
  p^{(m)} = q^{(m)}_{xx} - 2 \, u \ast q^{(m)} 
            - \sum_{k=0}^m q^{(k)} \ast q^{(m-k)}
            + \sum_{k=0}^{m-1} \left( q^{(k)} \ast q^{(m-1-k)} \right)_x
\ee
for $m \geq 1$. These formulas allow the recursive calculation of the functions 
$p^{(m)}$ and $q^{(m)}$ in terms of $u$ and its derivatives. From (\ref{w}) with
$w = \sum_{m \geq 0} w^{(m)}$ we now obtain the following expressions for the 
conserved densities, 
\be
    w^{(0)} &=& q^{(0)} = u  \, , \qquad  
    w^{(1)} = - u_x - \vartheta \, u \diamond 
                [ u_{xxx} - 3 \, (u \ast u)_x ]    \\
    w^{(m)} &=& q^{(m)} - \vartheta \, \sum_{k=0}^{m-1} q^{(k)} \diamond 
                p^{(m-1-k)}_x      \qquad (m \geq 1)  \; .
\ee

\section{From KdV solutions to ncKdV solutions}
\setcounter{equation}{0}
In this section we show that every solution of the KdV equation determines 
a solution of the ncKdV equation.
\vskip.1cm

Let $u' = \pa u/ \pa \vartheta$. Differentiation of (\ref{ncKdV}) with 
respect to $\vartheta$ leads to
\be
   {u'}_t + {u'}_{xxx} - 3 \, ( u' \ast u + u \ast u' )_x 
    - {3 \over 2} \, [ u_t , u_x ]_x = 0   \label{ncKdV'}
\ee
where
\be
    [ f , h ] = f \ast h - h \ast f  \; .
\ee
Using the identity
\be
  3 \, [ u_t , u_x ] = [ u , u_x]_t + [u , u_x ]_{xxx} 
                       - 3 \, \left( [ u , u_x ]_x \ast u
                       + u \ast [ u , u_x ]_x \right)
\ee
(\ref{ncKdV'}) can be rewritten as
\be
   z_t + z_{xxx} - 3 \, ( z \ast u + u \ast z )_x = 0  \label{z-eq}
\ee
where
\be
   z = u' - {1 \over 2} \, [ u , u_x ]_x 
     = u' - {1 \over 2} \, [ u , u_{xx} ]  \; .  \label{z-u'}
\ee
(\ref{z-eq}) is linear in $z$ and homogeneous. It admits the solution $z=0$, 
i.e.,
\be
    u' = {1 \over 2} \, [ u , u_{xx} ]  \; .  \label{z=0}
\ee 
Let us define
\be
    u_m =  \left. {\pa^m u \over \pa \vartheta^m} \right|_{\vartheta=0}  
           \qquad  (m \geq 0) \; .
\ee
\vskip.1cm
\noindent
{\em Lemma.} As a consequence of (\ref{z=0}) we have $u_{2m+1} = 0$ for 
all $m \geq 0$ and, for $m > 0$,
\be
      u_{2m}
  &=& \sum_{k=0}^{m-1} {1 \over 2^{2k+1}} \, {2m-1 \choose 2k+1} \, 
      \sum_{j=0}^{m-k-1} {2(m-k-1) \choose 2j} \sum_{i=0}^{2k+1} 
      {2k+1 \choose i} \nonumber \\
  & & (-1)^{2k-i+1} \left( \partial_t^i \pa_x^{2k-i+1} u_{2j} \right) 
      \left( \pa_t^{2k-i+1} \partial_x^{i+2} u_{2(m-k-j-1)} \right)
      \; .    \label{Lemma}
\ee
\vskip.1cm
\noindent
{\em Proof:} Using (\ref{z=0}) and $u \ast u_{xx} - u_{xx} \ast u 
  = 2 \, {\bf m} \circ \sinh(\vartheta P/2) \, (u \otimes u_{xx})$, one finds
\bea
    {\pa^{2m+1} u \over \pa \vartheta^{2m+1}} 
  = {\pa^{2m} \over \pa \vartheta^{2m}} {\bf m} \circ \sinh(\vartheta P/2) 
    \, (u \otimes u_{xx})
\eea
which implies
\bea
    u_{2m+1}
  = \sum_{k=0}^{m-1} {2m \choose 2k+1} \, 2^{-(2k+1)} \, {\bf m} \circ P^{2k+1}  
    \, \left. {\pa^{2(m-k)-1} \over \pa \vartheta^{2(m-k)-1}} (u \otimes u_{xx})
    \right|_{\vartheta=0} \; .
\eea
The summands on the rhs all have a factor consisting of an odd number of 
derivatives with respect to $\vartheta$ acting on $u \otimes u_{xx}$ according 
to the product rule of differentiation. This results in a sum of terms each of 
which has at least one odd derivative of $u$ or $u_{xx}$ as a factor. Using 
$u'|_{\vartheta=0} = 0$ which follows from (\ref{z=0}), our first assertion 
follows by induction. A similar calculation shows that, for $m > 0$,
\bea
  u_{2m} = \sum_{k=0}^{m-1} 2^{-(2k+1)} \, {2m-1 \choose 2k+1} 
           \sum_{j=0}^{m-k-1} {2(m-k-1) \choose 2j} \, {\bf m} \circ P^{2k+1} 
           \left( u_{2j} \otimes u_{2(m-k-j-1),xx} \right)
\eea 
taking $u_{2m+1}=0$ into account. Using the definition of $P$, we obtain the 
formula (\ref{Lemma}). 
     {  }   \hfill   \rule{5pt}{5pt}
\vskip.1cm

Now we have the following result: to every solution $u_0$ of the classical KdV 
equation there is a solution of the ncKdV equation at least as a formal power 
series in $\vartheta$,
\be
  u = u_0 + {1 \over 2} \, \vartheta^2 \, u_2 + {1 \over 4!} \, \vartheta^4 \, u_4
      + \ldots
\ee 
where the functions $u_{2m}$, $m>0$, have to be defined by (\ref{Lemma}).\footnote{The 
solution $u$ is real (if $u_0$ is real), irrespective of whether $\vartheta$ 
is real or imaginary. If $\vartheta$ is imaginary, then 
$\vartheta^{2m} = (-1)^m |\vartheta|^2$ and the power series is alternating.}
Hence, (\ref{z=0}) yields a transformation from the commutative KdV to the 
noncommutative model which is similar to the map considered by Seiberg and 
Witten in \cite{Seib+Witt99} (see section 3.1 there, in particular).\footnote{Of 
course, this map does not exhaust the set of solutions of the ncKdV equation since 
(\ref{z-eq}) also allows solutions with $z \neq 0$.} 
\vskip.1cm

Let us consider the one-soliton solution of the classical KdV equation,
\be
   u_0(x,t) = - 2 \, k^2 \, \mbox{sech}^2( k \, x - 4 \, k^3 \, t)
              \label{1s-KdV}
\ee
where $k$ is a constant. In this case also the ``even" coefficients 
(\ref{Lemma}) all vanish. For example, 
\be
   u_2 = {1 \over 2} \, ( u_{0,t} \, u_{0,xxx} - u_{0,x} \, u_{0,xxt} )
\ee
vanishes since $x$ and $t$ enter the one-soliton solution only 
through a single linear combination. One can also use an argument similar 
to that in \cite{DMH00d}, section 5, to verify that (\ref{1s-KdV}) 
is indeed an exact solution of the ncKdV equation. 
\vskip.1cm

\begin{figure}
\centering
\includegraphics{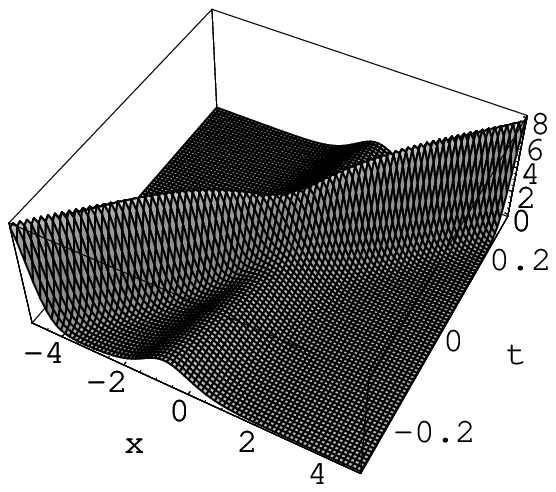} 
\caption{Plot of the two-soliton KdV solution. More precisely, 
the plot shows $-u_0$.}
\label{ncKdV0} 
\end{figure}

\begin{figure}
\centering
\includegraphics{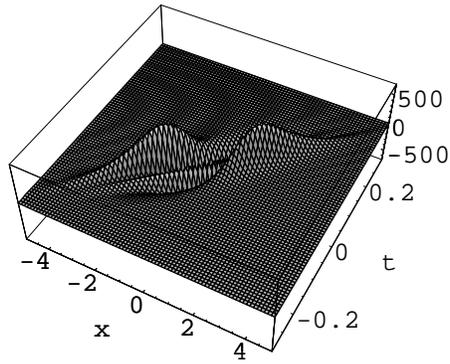} 
\caption{Plot of the second order ncKdV correction $u_2$ to the two-soliton 
KdV solution.}
\label{ncKdV2} 
\end{figure}

A two-soliton solution of the KdV equation is
\be
   u_0(x,t) = -12 \, {3 + 4 \, \cosh(2x-8t)+\cosh(4x-64t) \over
              [ 3 \, \cosh(x-28 t)+\cosh(3x-36t)]^2}
\ee
(see \cite{Draz+John89}, for example). In this case we get
\be
    u_2 = 331776 \, {( \cosh[3 \, x - 36 \, t] - 
            3 \, \cosh[x - 28 \, t] ) \, ( \sinh[3 \, x - 36 \, t] 
          + \sinh[x - 28 \, t])^2 \over ( \cosh[3 \, x - 36 \, t] 
          + 3 \, \cosh[x - 28 \, t] )^5 } 
\ee
with the help of Mathematica\footnote{Mathematica is a registered trademark 
of Wolfram Research. See also \cite{Math}.}.
In particular, the second order ncKdV correction to the above two-soliton KdV 
solution does not vanish. Plots of the two-soliton KdV solution and its 
ncKdV correction $u_2$, produced with Mathematica, 
are shown in Fig.~\ref{ncKdV0} and Fig.~\ref{ncKdV2}, respectively. 
\vskip.1cm

\begin{figure}
\centering
\includegraphics{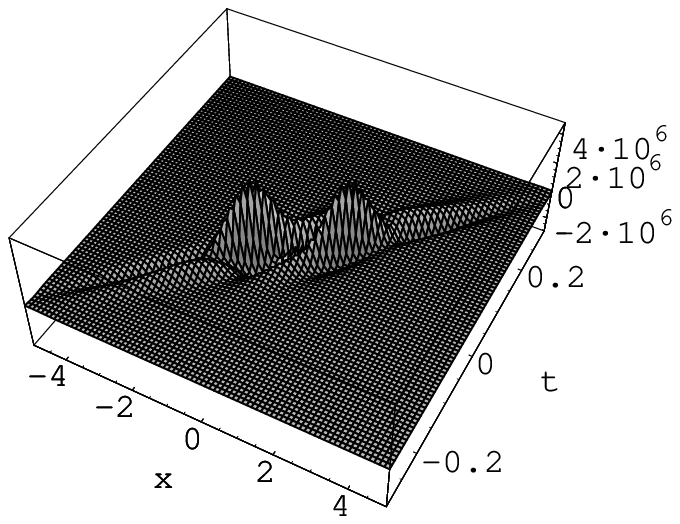} 
\caption{Plot of the forth order ncKdV correction $u_4$ to the two-soliton KdV 
solution.}
\label{ncKdV4} 
\end{figure}

 For the forth order ncKdV correction to the two-soliton KdV solution we get
\be
   u_4 &=& {3 \over 2} \, ( u_{2,t} \, u_{0,xxx} - u_{2,x} \, u_{0,txx}
           + u_{0,t} \, u_{2,xxx} - u_{0,x} \, u_{2,txx} ) \nonumber \\
       & & + {1 \over 8} \, ( u_{0,ttt} \, u_{0,xxxxx} 
           - 3 \, u_{0,ttx} \, u_{0,txxxx}
           + 3 \, u_{0,txx} \, u_{0,ttxxx} - u_{0,xxx} \, u_{0,tttxx} )
\ee
which results in a lengthy expression in terms of hyperbolic functions. 
This function is plotted in Fig. \ref{ncKdV4}. There is a strong similarity 
between $u_2$ and $u_4$. The plots show that $u_2$ and $u_4$ vanish as 
$t \to \pm \infty$. Indeed, since an $N$-soliton solution $u_0$ of the KdV 
equation asymptotically (as $t \to \pm \infty$) separates into single solitons, 
the ncKdV corrections $u_{2m}$, $m>0$, tend to zero by (\ref{Lemma}) and the 
properties of a single soliton solution.

\section{Noncommutative Miura transformation and noncommutative modified KdV equation}
\setcounter{equation}{0}
An obvious analogue of the classical {\em Miura transformation} in the noncommutative 
framework is
\be
    u = v_x + v \ast v  \; .
\ee
With its help one finds
\be
   \mbox{ncKdV}(u) = (\mbox{ncmKdV}(v))_x + v \ast (\mbox{ncmKdV}(v))
  + (\mbox{ncmKdV}(v)) \ast v
\ee
where $\mbox{ncKdV}(u)$ stands for the left hand side of (\ref{ncKdV}) and 
\be
  \mbox{ncmKdV}(v) = v_t + v_{xxx} - 3 \, (v \ast v \ast v_x + v_x \ast v \ast v)
                     \; .
\ee
Hence, if $v$ solves the {\em noncommutative modified KdV} (ncmKdV) 
equation 
\be
        \mbox{ncmKdV}(v) = 0   \label{ncmKdV}
\ee
then $u$ solves the ncKdV equation.
Differentiation of the Miura transformation with respect to $\vartheta$ leads to
\be
   u' = v'_x + v' \ast v + v \ast v' + {1 \over 2} \, [ v_t , v_x ] 
\ee
so that (\ref{z-u'}) translates to
\be
    v' = {1 \over 2} \, \left( [v \ast v , v_{xx} ] - [v , v_x \ast v_x] \right)
         + r     \label{v'-r}
\ee
where
\be
    z = r_x + r \ast v + v \ast r \; .
\ee
Differentiating (\ref{ncmKdV}) with respect to $\vartheta$, a lengthy calculation 
shows that the resulting equation is satisfied as a consequence of (\ref{v'-r}) with
$r=0$. This means that we have a construction of ncmKdV solutions from mKdV solutions 
in complete analogy to the ncKdV case treated in the previous section.
\vskip.1cm

Instead of the noncommutative Miura transformation we can consider the analogue 
of the Gardner transformation (cf \cite{Draz+John89})
\be
   u = q + \lambda \, q_x + \lambda^2 \, q \ast q
\ee
which is precisely our equation (\ref{q-eq}). Then one finds
\be
   \mbox{ncKdV}(u) = \left( 1 + \lambda \, \pa_x + \lambda^2 \, \{ q , \cdot \} \right)
                     \, \mbox{ncKdV}(q;\lambda)
\ee 
where $\{ f,h \} = f \ast h + h \ast f$ and
\be
   \mbox{ncKdV}(q;\lambda) = q_t + q_{xxx} - 3 \, (q \ast q)_x 
   - 3 \, \lambda^2 \, (q \ast q \ast q_x + q_x \ast q \ast q)  \; .
\ee
As a consequence, $u$ satisfies the ncKdV equation if $q$ is a solution of 
the noncommutative generalized KdV equation
\be
     \mbox{ncKdV}(q;\lambda) = 0  \; .
\ee
Using the identity
\be
    3 \, (q \ast q \ast q_x + q_x \ast q \ast q) 
  = 2 \, (q \ast q \ast q)_x + [q,[q,q_x]]
\ee
and (\ref{diamond}), the latter equation can be rewritten in the form of a 
conservation law as follows,
\be
    \tilde{w}_t
    + (q_{xx} - 3 \, q \ast q - 2 \, \lambda^2 \, q \ast q \ast q
    + \lambda^2 \vartheta \, q_t \diamond [q,q_x] )_x = 0
\ee
where
\be
    \tilde{w} = q - \lambda^2 \vartheta \, q_x \diamond [q,q_x] \; .
\ee
This expression is different from the conserved density $w$ given in (\ref{w}). 
However, $\tilde{w}$ and $w$ must be equal up to a total $x$-derivative.

\section{Conclusions}
\setcounter{equation}{0}
We obtained a (Moyal) deformed version of the KdV equation which 
lives on a noncommutative space-time and which shares with its classical 
version the property of having an infinite set of conserved densities, a 
characteristic feature of soliton equations and (infinite-dimensional) 
integrable models. Indeed, the soliton structure of the KdV equation is 
essentially preserved under the deformation so that the ncKdV equation is 
a very concrete example of a ``noncommutative soliton equation". 
More precisely, given a solution of the KdV equation, there is a 
prescription how to calculate from it a corresponding solution of the ncKdV 
equation order by order in $\vartheta$. In particular, the one-soliton 
solution of the KdV equation is also an exact solution of the ncKdV 
equation (without $\vartheta$-corrections). For the two-soliton solution, 
corrections of even order in the deformation parameter $\vartheta$ arise 
from the ncKdV equation which lead to modifications in the strong interaction 
region of the solitons. In principle, such modulations could be observed in a 
physical system. Furthermore, it is quite surprising that the relation 
between the KdV and the mKdV equation via the Miura transformation, as well 
as the construction of conserved densities via the Gardner transformation 
passes over to the noncommutative equations.

\newpage

\begin{thebibliography}{**}
\bibitem{Seib+Witt99} 
 N. Seiberg and E. Witten, {\em String theory and noncommutative geometry}, 
 JHEP 09 (1999) 032.
\bibitem{dq} F. Bayen, M. Flato, C. Fronsdal, A. Lichnerowicz and D. Sternheimer,
 {\em Deformation theory and quantization I, II}, Ann. Phys. 111 (1978) 61--151.
\bibitem{DMH00a} A. Dimakis and F. M\"uller-Hoissen, 
 {\em Bi-differential calculi and integrable models},  
 J. Phys. A: Math. Gen. 33 (2000) 957--974.
\bibitem{DMH00b} A. Dimakis and F. M\"uller-Hoissen, 
 {\em Bicomplexes and integrable models}, nlin.SI/0006029.
\bibitem{DMH99} A. Dimakis and F. M\"uller-Hoissen, 
 {\em Bi-differential calculus and the KdV equation}, 
 math-ph/9908016, to appear in Rep. Math. Phys..
\bibitem{DMH00c} A. Dimakis and F. M\"uller-Hoissen, 
 {\em Bicomplexes, integrable models, and noncommutative geometry}, hep-th/0006005. 
\bibitem{DMH00d} A. Dimakis and F. M\"uller-Hoissen, 
 {\em A noncommutative version of the nonlinear Schr\"odinger equation}, hep-th/0007015.
\bibitem{Stra97} I.A.B. Strachan, {\em A geometry of multidimensional integrable systems},  
 J. Geom. Phys. 21 (1997) 255--278.
\bibitem{Draz+John89} P.G. Drazin and R.S. Johnson, Solitons: an Introduction, 
 Cambridge University Press, Cambridge, 1989. 
\bibitem{Math} S. Wolfram, The Mathematica Book, 4th ed., Cambridge 
 University Press, Cambridge, 1999.
\end{thebibliography}
\end{document}